\documentclass[fleqn,usenatbib,useAMS]{mnras}

\usepackage[british]{babel}
\usepackage[T1]{fontenc}
\usepackage{ae,aecompl}

\usepackage{graphicx}	
\usepackage{amsmath}	
\usepackage{amssymb}	
\usepackage{xspace}
\usepackage{verbatim}

\newcommand{\galform}{{\sc{galform}}\xspace}
\newcommand{\mbh}{M_{\mathrm{BH}}}
\newcommand{\jbh}{J_{\mathrm{BH}}}

\newcommand{\ledd}{L_{\mathrm{Edd}}}

\title{The evolution of radio jets across cosmic time}

\author[A. J. Griffin et al.]{
Andrew J. Griffin,$^{1}$\thanks{E-mail: andrew.j.griffin@durham.ac.uk (AJG)}
Cedric G. Lacey,$^{1}$
Violeta Gonzalez-Perez,$^{1,2,3}$\newauthor
Claudia del P. Lagos$^{4,5,6}$
\\
\\
$^{1}$Institute for Computational Cosmology, Department of Physics, University of Durham, South Road, Durham, DH1 3LE, UK\\
$^{2}$Institute of Cosmology and Gravitation, University of Portsmouth, Burnaby Road, Portsmouth PO1 3FX, UK\\
$^{3}$ Energy Lancaster, Lancaster University, Lancaster, LA1 4YB, UK\\
$^{4}$International Centre for Radio Astronomy Research (ICRAR), M468, University of Western Australia, 35 Stirling Hwy, Crawley, \\  WA 6009, Australia\\
$^{5}$ARC Centre of Excellence for All Sky Astrophysics in 3 Dimensions (ASTRO 3D)\\
$^{6}$Cosmic Dawn Center (DAWN), Copenhagen, Denmark\\
}

\date{Accepted XXX. Received YYY; in original form ZZZ}

\pubyear{2019}

\begin{document}
\label{firstpage}
\pagerange{\pageref{firstpage}--\pageref{lastpage}}
\maketitle

\begin{abstract}
We present predictions for the evolution of radio emission from Active Galactic Nuclei (AGNs). We use a model that follows the evolution of Supermassive Black Hole (SMBH) masses and spins, within the latest version of the \galform semi-analytic model of galaxy formation. We use a Blandford-Znajek type model to calculate the power of the relativistic jets produced by black hole accretion discs, and a scaling model to calculate radio luminosities. First, we present the predicted evolution of the jet power distribution, finding that this is dominated by objects fuelled by hot halo accretion and an ADAF accretion state for jet powers above $10^{32}\mathrm{W}$ at $z=0$, with the contribution from objects fuelled by starbursts and in a thin disc accretion state being more important for lower jet powers at $z=0$ and at all jet powers at high redshifts ($z\geq3$). We then present the evolution of the jet power density from the model. The model is consistent with current observational estimates of jet powers from radio luminosities, once we allow for the significant uncertainties in these observational estimates.  Next, we calibrate the model for radio emission to a range of observational estimates of the $z=0$ radio luminosity function. We compare the evolution of the model radio luminosity function to observational estimates for $0<z<6$, finding that the predicted evolution is similar to that observed. Finally, we explore recalibrating the model to reproduce luminosity functions of core radio emission, finding that the model is in approximate agreement with the observations. 
\end{abstract}




\section{Introduction}

Active Galactic Nuclei (AGNs) play a crucial role in galaxy formation. Vast cavities are seen in X-ray images of the hot intracluster gas in clusters, which appear to be produced by the expansion of radio-emitting plasma produced by AGN activity \citep[e.g.][]{forman05,cavagnolo11,david11}. This AGN activity is also thought to heat the gas in haloes around individual galaxies, and so provides a mechanism for shutting off gas cooling in haloes and hence star formation in massive galaxies. This effect can produce the `red and dead' ellipticals observed in the contemporary Universe, and allows theoretical models to reproduce the bright end of the galaxy luminosity function at the present day \citep[e.g.][]{bower06, croton06a}. Given the role AGN feedback plays in the local Universe, understanding the evolution of AGN feedback through cosmic time is essential for an understanding of the evolution of the galaxy population. 

Understanding the cosmic evolution of extragalactic radio sources has been of interest to the astrophysical community since the 1960s, when the aim was to distinguish between a steady-state Universe, and an evolving Big Bang model. Early work showed that the most luminous radio sources exhibited stronger cosmological evolution than the less luminous sources \citep{longair66}, but the lack of radio source redshifts in that work constituted a major uncertainty. Subsequent work showed that the comoving number density of powerful radio sources at $z \sim 2$ is $\sim 1000$ higher than for the local Universe, with a strong decrease in the number density from $z=2$ to $z=4$ \citep[e.g.][]{peacock85,dunlop90}, which was referred to as the high redshift `cut-off'. Other works disputed this cut-off, with \cite{jarvis00}, \cite{jarvis01}, and \cite{willott01} suggesting a more gradual evolution in the number density of high-redshift sources. In a more detailed analysis, \cite{wall05} confirmed the decrease in the number density of flat-spectrum radio sources for $z \geq 3$. More recent studies have also investigated the less powerful sources, which seem to show only a modest increase in number density of a factor $\sim 2$ from $z=0$ to $z=0.5$ \citep[e.g.][]{sadler07,donoso09}, and also find that there is a decrease in number density for $z>0.7$ \citep[e.g.][]{rigby11}. Constraining the shape and evolution of the radio luminosity function is challenging observationally as it requires surveys with sufficient sensitivity and survey area to probe a wide range of radio luminosities.

This radio emission from AGNs is powered primarily by relativistic jets that originate near the SMBH. The most widely accepted models for these jets postulate that the jet energy source is either the rotational energy of the black hole \citep{blandfordznajek77}, or the rotational energy of the accretion disc \citep{blandfordpayne82}. Production of jets in these models also requires a strong magnetic field close to the black hole, which is assumed to be generated in a black hole accretion disc. The evolution of the radio population therefore depends on the evolution of black hole masses, spins and accretion rates.

The radio luminosity function of AGNs has been suggested to be composed of two evolving populations, namely of 
`radiative-mode' and `jet-mode' sources \cite[e.g.][]{bestheckman12}\footnote{Note that the radio emission of radiative mode sources is still thought to be powered by a jet.}. This classification is typically based on optical emission line strength.
Black holes in radiative-mode radio-AGN are thought to accrete material 
via a physically-thin, optically-thick, radiatively efficient accretion disc \citep{shakurasunyaev73}. 
Such AGNs have luminosities of $L \gtrsim 0.01\ledd$, where $\ledd$ is the Eddington luminosity,
and the radiation can also drive wide angle outflows at sub-relativistic velocities. On the other hand, black holes in jet-mode radio-AGN are thought to accrete material  
in a radiatively inefficient manner via a physically-thick, optically-thin accretion flow
dominated by advection, referred to as an Advection Dominated Accretion Flow \citep[ADAF - cf.][]{yn14}. These objects have luminosities of $L \lesssim 0.01 \ledd$, and most of the outflow energy is in collimated relativistic jets. 
\cite{hb14} provide a comprehensive review of these two modes from an observational perspective. 

Some studies have also investigated the evolution of the radio luminosity function for the compact central components (`cores') of radio AGNs. These typically use high frequency radio observations. The emission from the core is believed to more directly probe the region around the SMBH, and so the evolution of the core radio luminosity function may be more directly connected to the evolution of SMBHs across cosmic time.

This paper is one of a series of papers exploring SMBH and AGN properties within an existing theoretical model of galaxy formation. \cite{griffin19a}, building on the model of \cite{fani11}, presented a self-consistent model for the evolution of masses and spins of SMBHs within the semi-analytical galaxy formation model \galform \citep{cole00,lacey16}. They applied this model to investigate a wide range of SMBH and AGN properties that were compared to observations at $0 \leq z \leq 6$. In \cite{griffin19b} this model was used to make predictions for future surveys with JWST and other space-based telescopes at $z \geq 7$. In this paper, we explore jet powers and radio luminosities predicted by this model. While various previous theoretical studies have investigated the evolution of radio luminosities using different modelling techniques for the host galaxies, such as physical models of galaxy formation \citep[e.g.][]{fani11,hirschmann14}, or empirical galaxy evolution models \citep[e.g.][]{kaiser99,saxena17}, very few models base their radio luminosities on a self-consistent model for SMBH growth and spin evolution embedded in a physical model of galaxy formation. 


The outline of this paper is as follows. In Section \ref{sec:model} we describe the model used. In Section \ref{sec:jet} we present the predicted evolution of the jet powers. In Section \ref{sec:total_radio} we present the predicted total radio luminosity function evolution, and in Section \ref{sec:core_radio} we explore recalibrating the model to compare to the core radio luminosity function. In Section \ref{sec:conclusions} we present our conclusions.

\section{Model}
\label{sec:model}

\subsection{Galaxy formation model}

We calculate the evolution of the population of AGNs using the semi-analytic model of galaxy formation \galform. In \galform, which was first 
introduced in \cite{cole00}, 
a merger tree describing the formation history of each dark matter halo is populated with galaxies using analytic prescriptions for all of the baryonic physics in the form of a set of coupled differential equations 
that track the exchange of baryons between different galaxy components. 
Physical processes that are modelled include 
i) merging of dark matter haloes, 
ii) shock heating and radiative cooling of gas in haloes, 
iii) quiescent star formation in discs,
iv) starbursts, 
v) feedback from photoionisation, supernovae, and AGNs,  
vi) the chemical evolution of gas and stars, 
vii) galaxies merging inside haloes due to dynamical friction, 
viii) bar instabilities in galaxy discs, 
ix) the evolution of stellar populations, and 
x) the extinction and reprocessing of radiation by dust.
In this paper, we use the same model as in \cite{griffin19a}.
This uses the \galform model of \cite{lacey16}, as recalibrated for the Planck-Millennium simulation in \cite{baugh19}. The \cite{lacey16} model has been shown to match a broad range of observational 
data over a wide range of wavelengths and redshifts: from UV luminosity functions at $z=6$ to the K-band luminosity function at $z=0$, through to $850\mu$m number counts. 
The model here uses the P-Millennium dark matter simulation \citep{baugh19}, which assumes the \emph{Planck} cosmology \citep{planck14} (which is different from the \textit{Wilkinson Microwave Anisotropy Probe} (WMAP-7) cosmology \citep{komatsu11} assumed in \cite{lacey16}). This model uses a flat $\Lambda$CDM cosmology with $h=0.678$, $\Omega_{m}=0.307$, $\Omega_{b}=0.0483$, and $\sigma_{8}=0.829$. P-Millennium has a halo mass resolution of $2.12 \times 10^{9} h^{-1} M_{\odot}$ (corresponding to 20 dark matter particles) 
compared to $1.87 \times 10^{10} h^{-1} M_{\odot}$ for the dark matter simulation used in \cite{lacey16}. The resulting changes in a small number of \galform parameters are described in \cite{baugh19}. The properties of the SMBHs and AGNs in this model have been shown to be 
broadly consistent with observations \citep{griffin19a}.

\subsection{SMBH spin evolution}

The relativistic jets that power radio emission in AGNs are generally believed to arise from a combination of rotation and magnetic fields in the accretion disc around a black hole, with the source of the jet power believed to be either the rotational energy of the black hole \citep{blandfordznajek77}, or of the accretion disc \citep{blandfordpayne82}. Hence, from a model of SMBH mass accretion and spin evolution, we can make predictions for properties of radio jets.

The model for SMBH spin evolution used here is that presented in \cite{griffin19a}, 
which is an updated version of the \cite{fani11} model. 
While we refer the reader to \cite{griffin19a} for a detailed description of the model, we provide a brief
overview here. 

As gas in galaxies falls into the central regions, gas is accreted onto the SMBH, which carries angular momentum with it, therefore causing the SMBH to be spun up. However, it is not clear whether the direction of the angular momentum of the gas remains constant as it falls inwards, resulting in two different theoretical scenarios. In the `prolonged scenario' of SMBH accretion \citep{volonteri07}, the angular momentum of gas is assumed to remain in the same direction as the gas falls inwards, whereas in the `chaotic scenario' of SMBH accretion proposed by \cite{kph08} the angular momentum direction of the accreted gas is assumed to be periodically randomised. On even smaller scales, the angular momentum of the SMBH can be misaligned with the angular momentum of the gas disc, causing the SMBH to induce Lense-Thirring precession in the disc. This effect causes the disc to be warped on the smallest scales \citep{bardeenpetterson75}.

In our model, the angular momentum of the SMBH and the accretion disc are calculated analytically, and the model tracks the evolution of these angular mometa as the gas is accreted onto the SMBH. As in \cite{griffin19a} we adopt here the chaotic scenario of SMBH accretion, and following \cite{kph08}, we assume that the angular momentum direction of the gas is randomised every time a disc self-gravity mass of gas is consumed. As in \cite{griffin19a} we assume that the gas is consumed in increments of the self-gravity mass. 

The SMBH spin also changes when two SMBHs merge following a galaxy merger. The final spin depends on the spins of the two merging black holes, as well as the angular momentum of their binary orbit. To calculate the final spin, we use the expressions obtained from numerical simulations of BH-BH mergers. 

The SMBH spin is parametrised by the dimensionless spin parameter, $a=c J_{\mathrm{BH}} /G \mbh^2$, where $\mbh$ is the black hole mass and $\jbh$ is the angular momentum of the black hole.
$a=0$ signifies an SMBH that is not spinning and $a=1$ signifies an SMBH that is maximally spinning. 

\subsection{SMBH accretion rates}

\begin{table}
\centering
\caption{The values for the SMBH/AGN free parameters in the model. $\alpha_{\mathrm{cool}}$ the threshold of AGN feedback, $\epsilon_{\mathrm{heat}}$, the AGN heating efficiency, $f_{\mathrm{BH}}$, the fraction of the mass of stars formed in the starburst accreted onto the SMBH, $f_{\mathrm{Edd}}$, the maximum SMBH heating rate parameter, and $f_{\mathrm{q}}$, the ratio of the duration of the accretion event to the bulge dynamical timescale, have the same values as in \protect\cite{griffin19a}.}
\begin{tabular}{|c|c|}
\hline
Parameter & Value \\
\hline
$\alpha_{\mathrm{cool}}$ & 0.8 \\
\hline
$\epsilon_{\mathrm{heat}}$ & 0.02 \\
\hline
$f_{\mathrm{BH}}$ & 0.005 \\
\hline
$f_{\mathrm{Edd}}$ & 0.01 \\
\hline
$f_{\mathrm{q}}$ & 10 \\
\hline
\end{tabular}
\label{tab:free_params}
\end{table}

The jet power depends on the mass accretion rate onto the SMBH as well as its mass and spin. In the model, gas is made available for accretion onto the SMBH by two fuelling modes: either by quiescent accretion from gas in the hot gas atmospheres of massive haloes, or by starbursts triggered by galaxy mergers or disc instabilities. The SMBH mass accretion rate is calculated as in \cite{griffin19a}, where for starbursts, the mass of gas made available for accretion onto the SMBH, is a fixed fraction, $f_{\mathrm{BH}}$, of the mass of the stars formed in the starburst, $M_{\star, \mathrm{burst}}$. The value of $f_{\mathrm{BH}}$ adopted is given in Table \ref{tab:free_params} and is the same as in \cite{lacey16}. The gas is assumed to accrete onto the SMBH at a constant rate over a time $f_{\mathrm{q}}t_{\mathrm{bulge}}$, where $t_{\mathrm{bulge}}$ is the dynamical timescale of the bulge, 
and $f_{\mathrm{q}}$ is a free parameter which was calibrated in \cite{griffin19a}, and given in Table \ref{tab:free_params}. Hence the mass accretion rate for this mode is:

\begin{equation}
  \dot{M}_{\mathrm{acc}} = \frac{f_{\mathrm{BH}} M_{\star , \mathrm{burst}}}{f_{\mathrm{q}}t_{\mathrm{bulge}}}.
  \end{equation}
  
For hot halo accretion, which is the mode of SMBH accretion where AGN feedback is assumed to be active (see Section \ref{sec:AGNf} below), the mass accretion rate has the value required for the heating by the black hole to suppress gas cooling in the galaxy halo, and is given by:

\begin{equation}
  \dot{M}_{\mathrm{acc}} = \frac{L_{\mathrm{cool}}}{\epsilon_{\mathrm{heat}}  c^2}, \label{eq:hh_mdot}
\end{equation}

\noindent
where $L_{\mathrm{cool}}$ is the radiative cooling luminosity of the hot halo gas and $\epsilon_{\mathrm{heat}}$ is the assumed constant halo heating efficiency of the SMBH. The value of $\epsilon_{\mathrm{heat}}$ is given in Table \ref{tab:free_params}, and is the same as in \cite{lacey16}. If the SMBH is accreting via both the starburst mode and the hot halo mode, the mass accretion rate onto the SMBH is the sum of both modes.

From the mass accretion rate, we calculate a dimensionless mass accretion rate, $\dot{m}$, which we use when calculating jet powers and radio luminosities in Section \ref{sec:jet_radio_Ls}. $\dot{m}$ is defined as the SMBH mass accretion rate divided by the Eddington mass accretion rate:

\begin{equation}
 \dot{m} = \frac{\dot{M}_{\mathrm{acc}}}{\dot{M}_{\mathrm{Edd}}},
\end{equation}

\noindent where $\dot{M}_{\mathrm{Edd}}$ is defined as:

\begin{equation}
 \dot{M}_{\mathrm{Edd}} = \frac{\ledd}{0.1 c^2}.
\end{equation}

Note that we define the Eddington mass accretion rate for a nominal fixed accretion efficiency of 0.1, even though the actual radiative efficiency is assumed to depend on black hole spin as described in \cite{griffin19a}. The Eddington luminosity $\ledd$ is given by:

\begin{equation}
 \ledd = 1.26 \times 10^{38} \, \Big( \frac{\mbh}{M_{\odot}} \Big) \, \rm{ergs^{-1}},
\end{equation}

\noindent
where we are using the expression appropriate for pure hydrogen gas.

\subsection{AGN feedback in the model}
\label{sec:AGNf}

SMBHs release energy through gas accretion, resulting in AGN feedback. In \galform, we assume that the AGN feedback that occurs is \textit{radio mode} feedback \citep{bower06,croton06a}, in which energy released by gas accreting onto the SMBH from the hot gas halo powers a relativistic jet that deposits energy in the halo to balance radiative cooling. As in \cite{lacey16}, we assume that the heating from the jet balances radiative cooling if the following two conditions are met. First, the cooling time of the gas, $t_{\mathrm{cool}}$, needs to be sufficiently long compared to the free-fall time, $t_{\mathrm{ff}}$, as cooling needs to be in the quasi-hydrostatic cooling regime:

\begin{equation}
    t_{\mathrm{cool}}(r_{\mathrm{cool}}) / t_{\mathrm{ff}}(r_{\mathrm{cool}}) > 1 / \alpha_{\mathrm{cool}}, \label{eq:AGNf1}
\end{equation}

\noindent 
where $r_{\mathrm{cool}}$ is the cooling radius calculated using the procedure in \cite{lacey16} Section 3.3, and $\alpha_{\mathrm{cool}}$ is a free parameter, which was calibrated in \cite{lacey16}, and is given in Table \ref{tab:free_params}. Secondly, the cooling luminosity, $L_{\mathrm{cool}}$ needs to be below a fraction of the Eddington luminosity, as jet production for AGN feedback is assumed to occur only for SMBHs accreting at low Eddington accretion rates:

\begin{equation}
    L_{\mathrm{cool}} < f_{\mathrm{Edd}} L_{\mathrm{Edd}}, \label{eq:AGNf2}
\end{equation}

\noindent
where $f_{\mathrm{Edd}}$ is a free parameter. The adopted value is given in Table \ref{tab:free_params}. If these conditions are not met, then there is assumed to be no accretion from the hot halo, and no AGN feedback.

\subsection{Jet powers and radio luminosities}
\label{sec:jet_radio_Ls}

We calculate jet powers from black hole accretion discs following the \cite{blandfordznajek77} (BZ) model, in which the jet power is sourced from the rotational energy of the black hole. In the BZ model, the jet power, $Q$, depends on the black hole mass, the black hole spin, $a$, and the strength of the poloidal magnetic field, $B_{p}$:

\begin{equation}
    Q \propto B_{p}^2 M_{\mathrm{BH}}^2 \, \, a^2. 
\end{equation}

The poloidal magnetic field in the accretion disc is related to the azimuthal magnetic field strength, $B_{\phi}$, via $B_{p} \approx (H/R) B_{\phi}$, where $H/R$ is the ratio of disc half-thickness to the disc radius. For geometrically thick ADAFs, $H/R \sim 1$, whereas for (geometrically) thin discs, $H/R$ is given by the thin disc equations. In our model, the SMBH is in the ADAF regime for $\dot{m}<0.01$, and is in the thin disc regime for $\dot{m}>0.01$. The poloidal magnetic field can then be related to accretion disc quantities by assuming the magnetic field pressure is limited by the maximum pressure of the accretion disc \citep{moderski96}. This assumption of equipartition is likely to provide an upper limit on $B_{\phi}$. The jet powers (summed over both jets) are then given by the expressions in \cite{meier02}:

\begin{equation}
 Q_{\mathrm{ADAF}} = 2 \times 10^{45} \mathrm{ergs^{-1}} \Bigg( \frac{\mbh}{10^9 M_{\odot}} \Bigg) \Bigg( \frac{\dot{m}}{0.01} \Bigg) a^2, \label{eq:Q_adaf}
\end{equation}

\begin{equation}
 Q_{\mathrm{TD}} = 2.5 \times 10^{43} \mathrm{ergs^{-1}} \Bigg( \frac{\mbh}{10^9 M_{\odot}} \Bigg)^{1.1} \Bigg( \frac{\dot{m}}{0.01} \Bigg)^{1.2} a^2. \label{eq:Q_TD}
\end{equation}

The coefficient for the thin disc case is lower than for the ADAF case as a result of the smaller values of $H/R$ for thin discs, which reduce the poloidal magnetic field compared to the azimuthal magnetic field. 

To calculate the radio luminosity, $L_{\nu \mathrm{R}}$, at a particular frequency, we use the model of \cite{heinzsunyaev03}, which relates 
the radio luminosity of core-dominated sources to the black hole mass and mass accretion rate, using scaling relations based on physical arguments. 
This model does not include any explicit dependence on the black hole spin. It gives $L_{\nu \mathrm{R}} \propto (M_{\mathrm{BH}}\dot{m})^{1.42}$ for ADAFs, and $L_{\nu \mathrm{R}} \propto M_{\mathrm{BH}}^{1.42}$, for thin discs.
By combining these relations with those for the jet powers, we obtain expressions for the radio 
luminosities\footnote{Note that equations (\ref{eq:lr_adaf}) and (\ref{eq:lr_td}) are different to \cite{fani11} equations (44) and (45)}:

\begin{equation}
 \nu_{\mathrm{R}} L_{\nu \mathrm{R, ADAF}} = A_{\mathrm{ADAF}} \, Q_{\mathrm{ADAF}} \Bigg( \frac{\mbh}{10^9 M_{\odot}} \Bigg)^{0.42} \Bigg( \frac{\dot{m}}{0.01} \Bigg)^{0.42}, \label{eq:lr_adaf}
\end{equation}

\begin{equation}
 \nu_{\mathrm{R}} L_{\nu \mathrm{R, TD}} = A_{\mathrm{TD}} \, Q_{\mathrm{TD}} \Bigg( \frac{\mbh}{10^9 M_{\odot}} \Bigg)^{0.32} \Bigg( \frac{\dot{m}}{0.01} \Bigg)^{-1.2},
 \label{eq:lr_td}
\end{equation}

\noindent 
where $\nu_{\mathrm{R}}$ is the rest-frame frequency, and $A_{\mathrm{ADAF}}$ and $A_{\mathrm{TD}}$ are dimensionless free parameters of the \cite{heinzsunyaev03} model, as their scaling relations do not provide values for the normalisation coefficients. We allow $A_{\mathrm{ADAF}}$ and $A_{\mathrm{TD}}$ to vary independently, different from \cite{fani11} who required $A_{\mathrm{ADAF}} / A_{\mathrm{TD}}=100$. 

We choose the values of $A_{\mathrm{ADAF}}$ and $A_{\mathrm{TD}}$ to give the best agreement with the observed AGN radio luminosity 
function at $z=0$, as we show in Figure \ref{fig:L800_radio_lf_z0} in Section \ref{sec:total_radio} for our fiducial model. The values adopted for this study are given in Table \ref{tab:A_values}.
We assume a power law SED for the radio emission, $L_{\nu \mathrm{R}} \propto \nu^{-\alpha}$, with $\alpha=0.7$. This radio emission model in \galform with the parameters of the fiducial model has been used in \cite{izquierdo18} to make predictions for the environments of radio galaxies, and in \cite{amarantidis19} in a comparison of AGN luminosity functions from different theoretical models. The model for jet powers has also been used in \cite{ceraj18}, who compared it to their observational estimate of the evolution of the jet power density.

\subsection{AGN heating and jet efficiency}
\label{sec:efficiency_distinction}

In the hot halo mode of SMBH accretion, where AGN feedback is operational, the efficiency of SMBH heating of the halo gas is set to a constant value $\epsilon_{\mathrm{heat}}=0.02$ as in equation (\ref{eq:hh_mdot}), and Table \ref{tab:free_params}. This is the efficiency used in the galaxy formation model, which we will refer to as the \textit{AGN heating efficiency}. We also calculate jet powers from the SMBH spin, mass, and accretion rate in equations (\ref{eq:Q_adaf}) and (\ref{eq:Q_TD}), from which an alternative efficiency, $\epsilon_{\mathrm{jet}}= Q / (\dot{M} c^2)$, can be calculated, which we will refer to as the \textit{AGN jet efficiency}. If all of the energy in AGN jets were deposited in hot halo gas, one would expect $\epsilon_{\mathrm{heat}} = \epsilon_{\mathrm{jet}}$. We here present predictions for the evolution of jet powers and radio luminosities from an existing theoretical model of galaxy formation, and so in order to avoid modifying the underlying galaxy formation model, this condition was not imposed on $\epsilon_{\mathrm{heat}}$. Modifying the model so that it is self-consistent in this aspect is not entirely straightforward, and so we postpone this to a future paper. However, we do explore in Section \ref{sec:jet} whether the assumed AGN heating efficiency is similar to the average predicted AGN jet efficiency, finding that they are similar.

\section{Evolution of jet power density}
\label{sec:jet}

\begin{figure*}
\centering
\includegraphics[width=\linewidth]{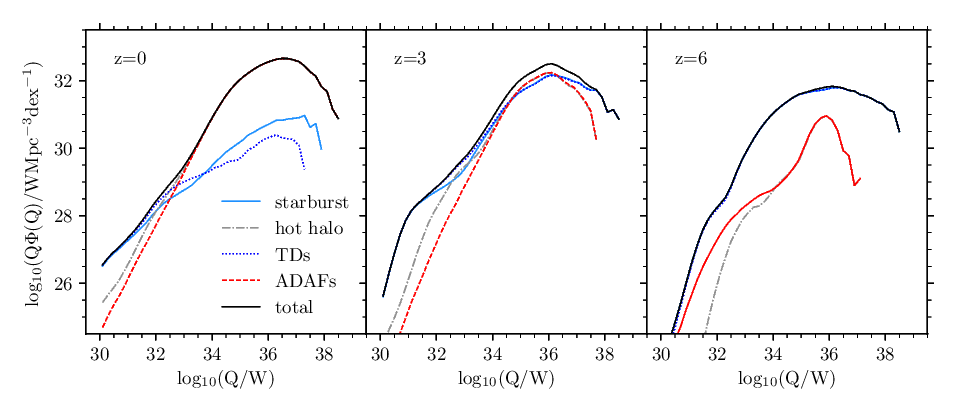}
\caption{The product of the jet power and the comoving number density of objects at each jet power $Q \Phi(Q) = Q dn / d \log Q$, as a function of jet power, $Q$, for $z=0,3,6$ (left, middle, and right panels). We show the total (black solid line), the total split by fuelling mode into the contribution from the starburst fuelling mode (light blue solid line) and from the hot halo fuelling mode (grey dot-dashed line), and the total split by accretion state into the contribution from thin discs (dark blue dotted line) and from ADAFs (red dashed line). When a source is fuelled by both the hot halo and starburst mode, we attribute the source to the mode with the highest mass accretion rate. In the left panel, the red and grey lines are underneath the black line, and in the right panel, the blue lines are underneath the black line.}
\label{fig:Q_distribution} 
\end{figure*}

\begin{figure}
\centering
\includegraphics[width=\linewidth]{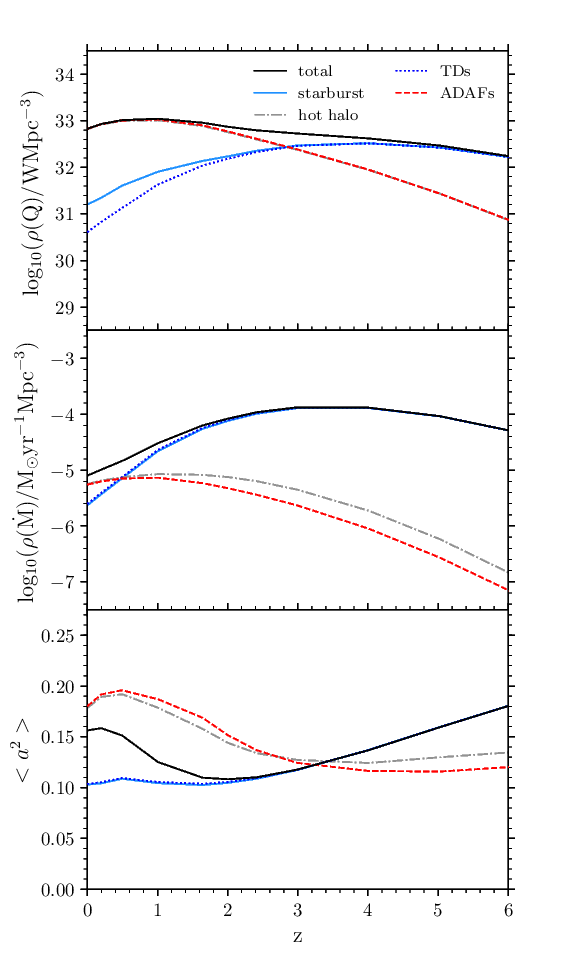}
\caption{The evolution of the model with redshift. In each panel the model prediction (black solid line),
is split into the contributions from starburst triggered accretion (light blue solid line), and from hot halo accretion (grey dot-dashed line). The model prediction is also split into the contributions from thin discs (TDs, dark blue dotted line) and from ADAFs (red dashed line). \emph{Top panel:} the predicted evolution of the jet power density with redshift. The solid grey line is underneath the red line. \emph{Middle panel:} the evolution of the SMBH mass accretion rate density with redshift. \emph{Bottom panel:} the evolution of the mean square SMBH spin (weighted by mass accretion rate) with redshift.}
\label{fig:Ljet_luminosity_density_Galform} 
\end{figure}

\begin{figure}
\centering
\includegraphics[width=\linewidth]{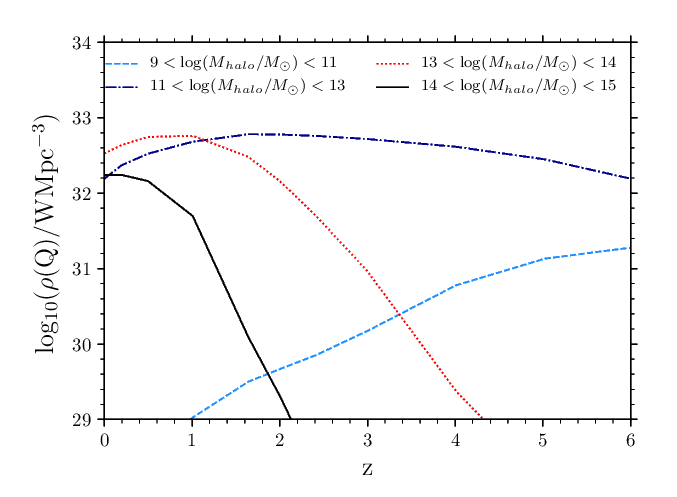}
\caption{The predicted evolution of the jet power density split into different bins in halo mass: $9 < \log (M_{\mathrm{halo}} / M_{\odot}) < 11$ (light blue dashed line), $11 < \log (M_{\mathrm{halo}} / M_{\odot}) < 13$ (dark blue dot-dashed line), $13 < \log (M_{\mathrm{halo}} / M_{\odot}) < 14$ (red dotted line), $14 < \log (M_{\mathrm{halo}} / M_{\odot}) < 15$ (black solid line).}
\label{fig:Ljet_luminosity_density_mhalo} 
\end{figure}

\begin{figure}
\centering
\includegraphics[width=\linewidth]{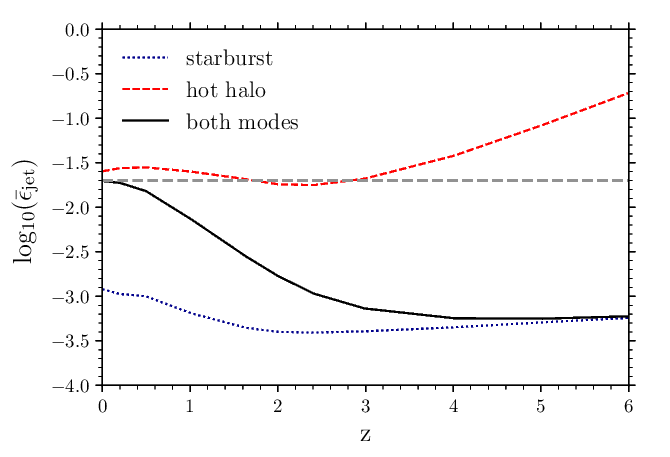}
\caption{The predicted evolution with redshift of the AGN jet efficiency, $\bar{\epsilon}_{\mathrm{jet}} = \rho (Q) / \rho(\dot{M}) c^2$ for the hot halo mode (red dashed line), for the starburst mode (blue dotted line), and for both modes combined (black solid line). We also show the assumed constant AGN heating efficiency of the galaxy formation model, $\epsilon_{\mathrm{heat}}=0.02$ (grey dashed line).}
\label{fig:epsilon_evolution} 
\end{figure}

\subsection{Predictions from the model}

We first investigate the predicted evolution of the jet powers. In Figure \ref{fig:Q_distribution}, we show $Q \Phi(Q)$, the product of the jet power and the comoving number density of objects at each jet power, $\Phi(Q) = dn / d \log Q$, for $z=0,3,6$. This is shown split by accretion state into the contributions from thin discs and ADAFs, and separately by fuelling mode into contributions from starburst-triggered accretion and hot halo accretion. When objects are fuelled by both the hot halo and the starburst modes, the source is attributed in the plot to whichever mode has the higher mass accretion rate. This distribution shows which jet powers and which contributions dominate the jet power density, as the integral of $Q \Phi(Q)$ with respect to $\log Q$ is equal to the jet power density. Note that our predictions for jet powers are independent of our model for radio emission. 

The ADAF and hot halo mode contributions evolve similarly because the Eddington normalised mass accretion rate for objects in the hot halo mode is generally below 0.01 (cf. equation (\ref{eq:AGNf2})). On the other hand, the thin disc and starburst mode contributions evolve similarly because the Eddington normalised mass accretion rate is generally above 0.01 for starburst mode accretion. This is because the mass accretion rate is typically higher for starburst mode accretion, and because the starburst mode typically occurs for smaller black holes in smaller haloes. At $z=0$, for $Q \lesssim 10^{32}~\mathrm{W}$, the dominant contribution to $Q\Phi(Q)$ is from the starburst and thin disc contributions, whereas for $Q \gtrsim 10^{33}~\mathrm{W}$, the dominant contribution is from the hot halo mode and ADAF contributions. At $z=3$, for $Q \lesssim 10^{34}~\mathrm{W}$, the starburst and thin disc contributions dominate, whereas at $Q \sim 10^{34}-10^{36}~\mathrm{W}$, the contributions to $Q\Phi(Q)$ from the starburst and hot halo modes contribute approximately equally. At $z=6$, the dominant contribution to $Q\Phi(Q)$ at all jet powers is from the starburst mode and thin disc contributions. These predictions for the different contributions to the jet power distribution could be tested observationally by future surveys.  

The jet power density discussed below is dominated by objects in the peak of the $Q \Phi(Q)$ distribution. The dominant contribution to the jet power density comes from sources with $Q \sim 10^{36}~\mathrm{W}$, independent of redshift over the range $0<z<6$. Also, the peak in $Q \Phi(Q)$ occurs at roughly the same jet power for both starburst and hot halo modes, again roughly independent of redshift. This appears to be fortuitous, given the very different typical jet efficiencies for the starburst and hot halo modes as discussed below. 

In the top panel of Figure \ref{fig:Ljet_luminosity_density_Galform} we show the evolution with redshift of the jet power density, $\rho(Q)$, where $\rho(Q)$ is given by the total jet power summed over all galaxies divided by the total comoving volume. We also split the jet power density evolution into contributions from thin discs and ADAFs, and separately into contributions from starburst triggered accretion and hot halo accretion. When comparing the fuelling modes, the hot halo mode contribution dominates the jet power density for $z<3$, whereas the starburst mode contribution dominates for $z>3$. The hot halo contribution peaks at $z \sim 1$, whereas the starburst contribution peaks at $z \sim 4$. When comparing the accretion disc states, the ADAF contribution dominates for $z<3$, and the thin disc contribution dominates for $z>3$.

In the middle panel of Figure \ref{fig:Ljet_luminosity_density_Galform}, we show the evolution with redshift of the SMBH mass accretion rate density (the total mass accretion rate summed over all galaxies divided by the total comoving volume). The total mass accretion rate density increases with redshift for $0<z<3$, has a peak around $z=3-4$, and then decreases for $z>4$. The mass accretion rate density is dominated by the contributions from AGNs fuelled by the starburst mode and accreting via the thin disc accretion state, except for $z<0.5$ where the mass accretion rate density is dominated by the contribution from AGNs fuelled by the hot halo mode and accreting via the ADAF accretion state. 

In the bottom panel of Figure \ref{fig:Ljet_luminosity_density_Galform}, we show the mass accretion rate weighted mean square SMBH spin, $\langle a^2 \rangle$, calculated as a ratio of densities, $\langle a^2 \rangle = \rho(\dot{M}a^2) / \rho(\dot{M})$. When considering all SMBHs together, $\langle a^2 \rangle$ decreases with redshift in the interval $0<z<2$, from about 0.15 to 0.1, before increasing for $z>2$ to about 0.18 at $z=6$. For $z<4$, the hot halo and ADAF contributions have higher values of $\langle a^2 \rangle$ compared to the starburst and thin disc contributions. This is because for the AGNs fuelled by the starburst mode, the objects with the highest mass accretion rates have low spins (around $a=0.2$ at $z=0$), whereas for the hot halo mode, the objects with the highest mass accretion rates have slightly higher spins ($a=0.2-0.4$ at $z=0$). For $z>4$, $\langle a^2 \rangle$ is greater for the starburst mode and thin disc contributions because the highest mass accretion rate SMBHs in the starburst mode have higher spins ($a=0.2-0.5$ at $z=6$), compared to the hot halo mode (where $a=0.2-0.4$ at $z=6$). The SMBH spin distributions are described in \cite{griffin19a}. 

By comparing Figure \ref{fig:Ljet_luminosity_density_Galform} to the expressions for the jet power in equations (\ref{eq:Q_adaf}) and (\ref{eq:Q_TD}), we see that the dominance of the hot halo contribution to the jet power density at $z<3$ is mainly due to the 80 times larger normalisation coefficient for ADAFs compared to thin discs. The relative evolution of the jet power densities from the starburst and hot halo modes is therefore driven mainly by the differences in mass accretion rates and in the normalisations of the jet power relations (see equations (\ref{eq:Q_adaf}) and (\ref{eq:Q_TD})), with variations in the spin playing only a minor role. 

In Figure \ref{fig:Ljet_luminosity_density_mhalo}, we present the jet power density split into the contribution from different halo masses. For $z<1$, the jet power density is dominated by AGNs in haloes of mass $13 < \log (M_{\mathrm{halo}} / M_{\odot}) < 14$ (i.e. large galaxy groups and clusters), whereas for $z>1$, the jet power density is dominated by AGNs in haloes of mass $11 < \log (M_{\mathrm{halo}} / M_{\odot}) < 13$ (i.e. individual galaxies and smaller groups). 

In Figure \ref{fig:epsilon_evolution}, we show the evolution of the mean AGN jet efficiency, $\bar{\epsilon}_{\mathrm{jet}}$ (cf. Section \ref{sec:efficiency_distinction}), which is calculated as the ratio of the jet power density to the mass accretion rate density. The mean AGN jet efficiency is higher for the hot halo mode than for the starburst mode at all redshifts: this is mainly because the normalisation coefficient of the jet power for ADAFs is higher than for thin discs by a factor of 80 (see equations (\ref{eq:Q_adaf}) and (\ref{eq:Q_TD})). The jet efficiency of the two modes combined is similar to the hot halo mode jet efficiency for $z<1$, and similar to that of the starburst mode for $z>3$.

For the hot halo mode, which is where AGN feedback is assumed to be active in the model, at $z=0$ the AGN jet efficiency is $\bar{\epsilon}_{\mathrm{jet}} \approx 0.03$, whereas at $z=3$ it is $\bar{\epsilon}_{\mathrm{jet}} \approx 0.02$, and at $z=6$ it is $\bar{\epsilon}_{\mathrm{jet}} \approx 0.2$, with an average over the history of the universe of $0.025$. This time averaged value of the AGN jet efficiency is only $25 \%$ larger than the assumed constant value of the AGN heating efficiency, $\epsilon_{\mathrm{heat}}=0.02$. The fact that the mean AGN jet efficiency for the hot halo mode only varies modestly with time suggests that the assumption that the AGN heating efficiency is constant through time is a reasonable approximation.

\subsection{Comparison of jet power density to observational estimates}

\begin{figure}
\centering
\includegraphics[width=\linewidth]{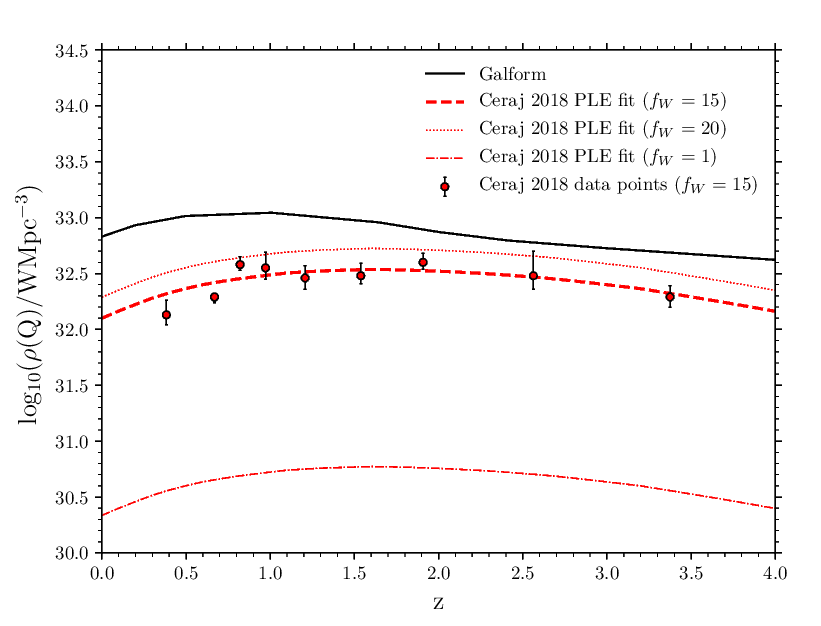}
\caption{The predicted evolution with redshift of the jet power density in the model compared to the observational estimate from \protect\cite{ceraj18}. 
The model prediction (solid black line),
is compared to the observational estimate for $f_{W}=15$, in redshift bins (red circles), and also using their pure luminosity evolution (PLE) fit to the data (dashed red line). We also show the estimates for the jet power density evolution from \protect\cite{ceraj18} for $f_{W}=1$ (dot-dashed red line) and $f_{W}=20$ (dotted red line). We only compare to the observational fit for $z \leq 4$, as at higher redshifts the fit is not well constrained by the data.}
\label{fig:Ljet_luminosity_density_obs_comp} 
\end{figure}

\begin{figure}
\centering
\includegraphics[width=\linewidth]{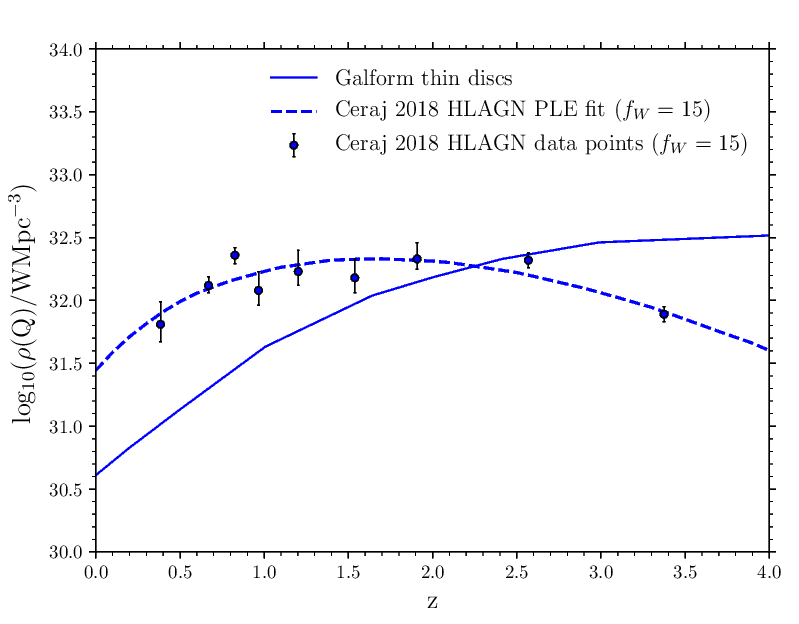}
\caption{The predicted evolution of the jet power density from thin discs in the model (blue solid line) compared to the evolution of the jet power of `HLAGN' from \protect\cite{ceraj18} (the dashed blue line is the pure luminosity evolution (PLE) fit to the data, and the blue points are the data in redshift bins).}
\label{fig:Ljet_luminosity_density_obs_comp_HLAGN} 
\end{figure}

We now compare the jet power density evolution predicted by our model to the observational estimate of \cite{ceraj18}. They obtain their estimate by measuring the evolution of the radio luminosity function at 1.4GHz, converting the 1.4GHz radio luminosities to jet 
powers using the \cite{willott99} relation, and then integrating over the subsequent jet power distribution. \cite{ceraj18} present their results both as data points in redshift bins, based on fitting an analytical luminosity function to data at that redshift, and also as a smooth function of redshift, obtained from an analytical pure luminosity evolution (PLE) model fit to their radio data.

The \cite{willott99} relation for jet power is derived using minimum energy arguments to estimate the energy stored in the lobes given the observed synchrotron luminosity and combining with an estimate of the source age based on a dynamical model for the lobe expansion. This relation is expressed in terms of 1.4GHz luminosity in \cite{hb14}, and is given by:

\begin{equation}
    Q = 4 \times 10^{35}\mathrm{W} \Bigg( \frac{L_{\nu R,  \mathrm{1.4GHz}}}{10^{25} \mathrm{WHz^{-1}}} \Bigg)^{6/7} (f_{\mathrm{W}})^{3/2},
\end{equation}

\noindent
where $f_{\mathrm{W}}$ is a factor that accounts for uncertainties in the knowledge of the physics of radio sources (primarily the composition of the radio emitting plasma and the low energy cutoff of the electron energy distribution). \cite{willott99} estimate $f_{\mathrm{W}}$ to lie in the range $f_{\mathrm{W}}=1-20$. Observational studies based on cavities in X-ray emitting hot gas around galaxies calculate the jet power from cavity volumes and pressures, and an estimate of the lifetime of the cavity based on the buoyancy timescale \citep{rafferty06,birzan08,cavagnolo10}. \cite{hb14} compiled these observational estimates of the jet power versus radio luminosity to find that they are consistent with the Willott et al. relation, with $f_{\mathrm{W}}=15$. Using a different method based on estimating lobe expansion velocities using spectral ageing, \cite{daly12} also find radio luminosities and jet powers consistent with the Willott et al. relation, for a value of $f_{\mathrm{W}}=4$. Other studies argue that other variables need to be considered in this relation, such as lobe size (because of radiative losses by the electron populations) \citep[e.g.][]{shabalagodfrey13}, the environment of sources \citep[e.g.][]{hardcastle18}, and Fanaroff-Riley morphology \citep[e.g.][]{turner15}.

We compare our predicted jet power density to the observational estimate of \cite{ceraj18}, in Figure \ref{fig:Ljet_luminosity_density_obs_comp}. Comparing to their observational estimate using a value of $f_{\mathrm{W}}=15$, our model is above their estimate by a factor of about 2 for $2 < z < 4$, and by a factor of about 4 for $z<2$. On the other hand, comparing to their observational estimate using a value of $f_{\mathrm{W}}=20$, our model is above their estimate by a factor of about 1.5 for $2 < z < 4$, and by a factor of about 2.5 for $z<2$. The jet power density in the model generally evolves in a similar way to the observations, 
with both the model and observations showing an increase in jet 
power density with redshift for $z \lesssim 1$, and a decrease for $z \gtrsim 1$. However, the increase of the jet 
power density with redshift for $z \lesssim 1$ is slightly less steep in the model compared to the observations, and the model evolution is also slightly less steep compared to the observations for $z \gtrsim 3$. 

While the model appears to be in some modest tension with the observations, there are several uncertainties in the observations to consider. First, there is uncertainty in the mean value of $f_{\mathrm{W}}$, which is estimated to take values in the range $1-20$ \citep{willott99}, and given that jet power depends on $f_{\mathrm{W}}$ as $Q \propto f_{\mathrm{W}}^{3/2}$, there are then significant uncertainties in the calculated $Q$ values. Secondly, rather than each radio source having the same $f_{\mathrm{W}}$ value, it is likely that the radio source population has a distribution of $f_{\mathrm{W}}$ values around the mean, resulting in an increase in the derived jet power density due to the non-linear dependence of $Q$ on $f_{\mathrm{W}}$. Thirdly, the observational estimate extrapolates the fit of the luminosity function to lower radio luminosities when integrating over luminosity to derive jet power densities. The faint end of the radio luminosity function may behave differently from the fit, which would affect the estimated jet power density. Overall, given the uncertainties, the model is reasonably consistent with the observations.

In Figure \ref{fig:Ljet_luminosity_density_obs_comp_HLAGN}, we compare the predicted jet power density for AGNs accreting via a thin disc accretion state to the jet power density of `moderate-to-high radiative luminosity AGN' (HLAGN) estimated from the observations by \cite{ceraj18}. In the observations, HLAGN are selected using a combination of (i) a threshold X-ray luminosity, (ii) mid-infrared colour-colour selection and (iii) template fitting to the spectral energy distributions \citep[see][]{ceraj18}. While HLAGN in the observations do not exactly correspond to thin discs in the model, one would expect radio sources accreting via a thin disc accretion state to have relatively high radiative luminosities, and so this is an approximate comparison. We find that while the model underpredicts the observations for $z<3$ and overpredicts them for $z>3$, it reproduces the behaviour that the number density of these objects should increase with redshift for $0<z<2$. A more detailed application of these selections for HLAGN to the model in the future might give closer agreement with the observations. Similar comparisons to Figures \ref{fig:Ljet_luminosity_density_obs_comp} and \ref{fig:Ljet_luminosity_density_obs_comp_HLAGN} were done in \cite{ceraj18}, based on a slightly earlier version of the model.

\section{Evolution of the total radio luminosity function of AGN}
\label{sec:total_radio}

\begin{figure*}
\centering
\includegraphics[width=\linewidth]{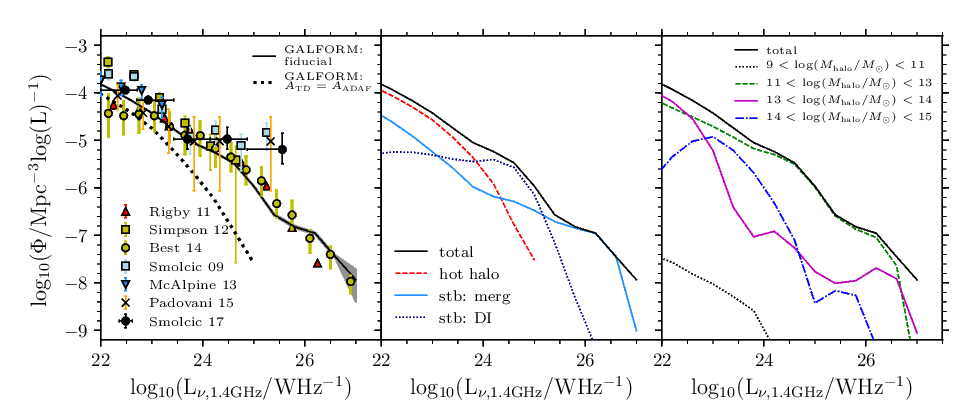}
\caption{\emph{Left panel:} the 1.4GHz luminosity function at $z~=~0$ from the model compared to observational estimates from VLA-COSMOS \citep[][light blue squares]{smolcic09}, CENSORS \citep[][red triangles]{rigby11}, 
the Subaru/XMM-Newton Deep Field radio source sample \citep[][yellow squares]{simpson12}, another VLA survey \citep[][blue triangles]{mcalpine13},
a sample from combining eight different surveys \citep[][yellow circles]{best14}, the Extended CDF South VLA sample \citep[][crosses]{padovani15} and COSMOS 3GHz data \citep[][black circles]{smolcic17}. The solid line is the prediction from the fiducial model, with the shaded region representing the Poisson errorbars. We also show the prediction of the model if we force $A_{\mathrm{TD}}$ to have the same value as $A_{\mathrm{ADAF}}$, where the value of $A_{\mathrm{ADAF}}$ is as in the fiducial model (dotted line). \emph{Middle panel:} the predicted radio luminosity function at $z=0$ (black solid line) split into contributions from the hot halo mode (red dashed line), starbursts triggered by mergers (light blue solid line), and starbursts triggered by disc instabilities (dark blue dotted line). \emph{Right panel:} the the predicted radio luminosity function at $z=0$ (black solid line) split into the contributions from haloes of mass $9 < \log(M_{\mathrm{halo}}/M_{\odot}) < 11$ (black dotted line), $11 < \log(M_{\mathrm{halo}}/M_{\odot}) < 13$ (green dashed line), $13 < \log(M_{\mathrm{halo}}/M_{\odot}) < 14$ (purple solid line), and $14 < \log(M_{\mathrm{halo}}/M_{\odot}) < 15$ (blue dot-dashed line).}
\label{fig:L800_radio_lf_z0} 
\end{figure*}

\begin{figure*}
\centering
\includegraphics[width=\linewidth]{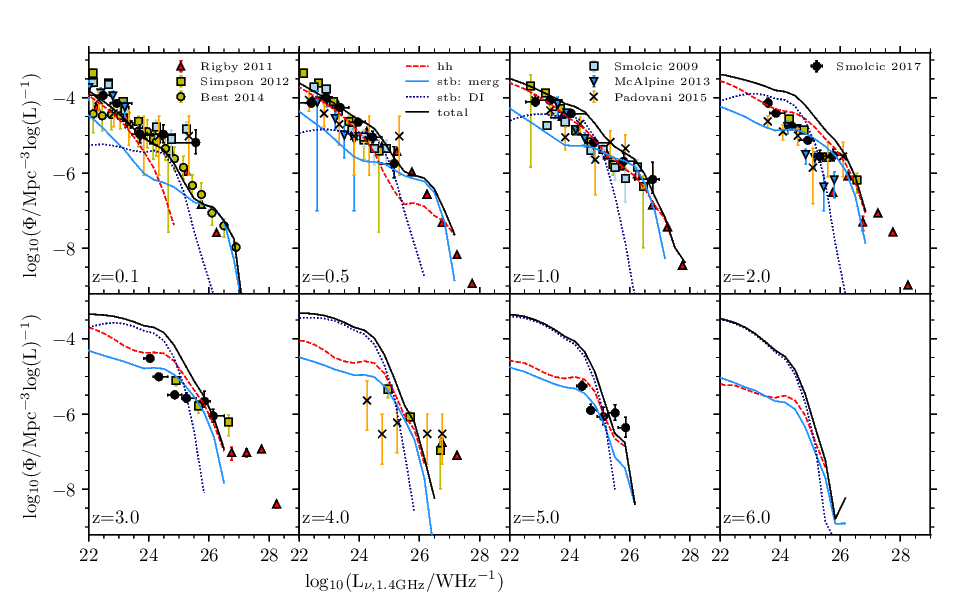}
\caption{The evolution of the predicted rest-frame 1.4GHz luminosity function of AGN compared to observational estimates. The symbols for the observations, and the linestyles for the different fuelling modes of the model are the same as for the middle panel of Figure \ref{fig:L800_radio_lf_z0}.}
\label{fig:L800_radio_lf_evolution} 
\end{figure*}

\begin{table}
\centering
\caption{The values of $A_{\mathrm{ADAF}}$ and $A_{\mathrm{TD}}$ for the different models presented in this paper. $A_{\mathrm{ADAF}}$ and $A_{\mathrm{TD}}$ are the normalisation coefficients of the radio luminosity for ADAFs and thin discs.}
\begin{tabular}{|c|c|c|}
\hline
Model & $A_{\mathrm{ADAF}}$ & $A_{\mathrm{TD}}$ \\
\hline
Fiducial & $2 \times 10^{-5}$ & $0.8$ \\
\hline
CR1 & $3 \times 10^{-5}$ & $3 \times 10^{-5}$ \\
\hline
CR2 & $3 \times 10^{-6}$ & $0.1$ \\
\hline
\end{tabular}
\label{tab:A_values}
\end{table}

We now compare our model predictions for the total radio luminosity function of AGN at $z=0$ with observational estimates, before analysing the evolution of the radio luminosity function. In this Section, we present results from the fiducial model, with the values of $A_{\mathrm{ADAF}}$ and $A_{\mathrm{TD}}$ given in Table \ref{tab:A_values}.
In the left panel of Figure \ref{fig:L800_radio_lf_z0} we show the 1.4GHz radio luminosity function predicted by the fiducial model at $z=0$ compared to observational estimates of the total radio luminosity function of AGN (i.e. including both compact and extended radio emission). The values of $A_{\mathrm{ADAF}}$ and $A_{\mathrm{TD}}$ were calibrated to the data shown in this figure. We allowed the values of $A_{\mathrm{ADAF}}$ and $A_{\mathrm{TD}}$ to be different. $A_{\mathrm{ADAF}}$ is primarily constrained by the faint end of the radio luminosity function ($L_{\nu} \sim 10^{22}\mathrm{WHz^{-1}}$), whereas $A_{\mathrm{TD}}$ is primarily constrained by the bright end of the radio luminosity function ($L_{\nu} \sim 10^{26}\mathrm{WHz^{-1}}$). $A_{\mathrm{TD}}$ needs to be larger than $A_{\mathrm{ADAF}}$ for the model to be in agreement with the bright end of the radio luminosity function. If we force $A_{\mathrm{TD}}$ to have the same value as $A_{\mathrm{ADAF}}$, the radio luminosity function predicted by the model is much steeper than the observations, as we show in the left panel of Figure \ref{fig:L800_radio_lf_z0}.

The model is able to match the observations very well at $z=0$, although there are some tensions between the different observational data sets, which we now describe. First, at $L_{\nu} \sim 10^{25} \mathrm{WHz^{-1}}$, the \cite{smolcic09} and \cite{smolcic17} number densities are about 10 times higher than 
the \cite{rigby11} and \cite{best14} number densities.
As discussed in Section 6.1.1 of \cite{padovani15}, this discrepancy may be a result of the sample selection. \cite{rigby11} and \cite{best14} select a sample of steep-spectrum sources 
($\alpha>0.5$)\footnote{This is for an assumed radio spectrum $S_{\nu} \propto \nu^{-\alpha}$}, in a variety of surveys with successively smaller areas and smaller flux density limits, whereas \cite{smolcic09} and \cite{smolcic17} select their sample with only a flux density limit. This difference could also be caused by sample variance caused by large scale structure, with the volumes at $z \sim 0$ probed by the surveys in \cite{smolcic09} and \cite{smolcic17} being relatively small. The quoted observational errors at this luminosity are fairly large, with the errors for \cite{smolcic09}, \cite{best14}, and \cite{smolcic17} being about 0.5 dex. Our predictions follow the estimate by \cite{rigby11} and \cite{best14} in this regime and to higher luminosities.

Secondly, for $L_{\nu} < 10^{23} \mathrm{WHz^{-1}}$, 
there is variation among the observational estimates spanning a range of around $\sim 1$ dex, which may also be for the same reason as for $L_{\nu} \sim 10^{25} \mathrm{WHz}^{-1}$. Alternatively, the differences may be caused by how the observational estimates distinguish between radio emission from star formation and from AGNs. \cite{pracy16} find that below $L_{\nu} \sim 10^{23} \mathrm{WHz^{-1}}$, the contribution from star formation dominates the radio luminosity function. Our model follows the data from \cite{mcalpine13} and \cite{smolcic17} most closely. These two regimes may warrant further observational studies to better constrain the radio luminosity function of AGNs at these luminosities.

In the middle panel of Figure \ref{fig:L800_radio_lf_z0} we present the radio luminosity function of the model for AGN at $z=0$ split by fuelling mode into contributions from the hot halo mode and from starbursts triggered by mergers and disc instabilities. The contribution from the hot halo mode is dominant for $L_{\nu} < 10^{24} \mathrm{WHz^{-1}}$, while for $10^{24} \mathrm{WHz^{-1}} < L_{\nu} < 10^{26} \mathrm{WHz^{-1}}$ the dominant contribution is from starbursts triggered by disc instabilties, and for $ L_{\nu} > 10^{26} \mathrm{WHz^{-1}}$ the dominant contribution is from starbursts triggered by galaxy mergers.

In the right panel of Figure \ref{fig:L800_radio_lf_z0} we show the radio luminosity function of the model at $z=0$ for AGN split into contributions from AGNs in different mass haloes. We find that for $10^{22.5} \mathrm{WHz^{-1}} < L_{\nu} < 10^{26.5} \mathrm{WHz^{-1}}$, the contribution from haloes of mass $11 < \log(M_{\mathrm{halo}}/M_{\odot}) < 13$ (individual galaxies and smaller groups) dominates, whereas for $L_{\nu} < 10^{22.5} \mathrm{WHz^{-1}}$, and $L_{\nu} > 10^{26.5} \mathrm{WHz^{-1}}$ the contribution from AGNs in haloes of mass $13 < \log(M_{\mathrm{halo}}/M_{\odot}) < 14$ (large galaxy groups and clusters) dominates. The $z=0$ radio luminosity function at intermediate luminosities ($10^{22.5} \mathrm{WHz^{-1}} < L_{\nu} < 10^{26.5} \mathrm{WHz^{-1}}$) is therefore predicted to probe AGNs in different mass haloes to the jet power density at $z=0$, where the latter is dominated by AGNs in haloes of mass $13 < \log(M_{\mathrm{halo}}/M_{\odot}) < 14$. 

In Figure \ref{fig:L800_radio_lf_evolution} we present the predicted evolution of the AGN radio luminosity function for 
$0<z<6$ compared to observational estimates.
The model prediction fits well to the observations at $z=0$ as previously discussed, but evolves differently compared to the observations. 
At $z=1$, for $L_{\nu} > 10^{26} \mathrm{WHz^{-1}}$ and for $L_{\nu} < 10^{24} \mathrm{WHz^{-1}}$, the model prediction is still in good agreement with the observations. However, the model overpredicts the number of objects around $L_{\nu} \sim 10^{25} \mathrm{WHz^{-1}}$ by about 0.5 dex. As we look to redshifts $z>3$, a trend emerges where the model 
overpredicts the luminosity function for low luminosities, underpredicts the luminosity function for high luminosities,
but predicts a similar number density to the observations for intermediate luminosities. 
For example, at $z=4$, the model overpredicts the number density for $L_{\nu} < 10^{25} \mathrm{WHz^{-1}}$,
underpredicts the number density for $L_{\nu} > 10^{26} \mathrm{WHz^{-1}}$, and agrees with the observations for $10^{25} \mathrm{WHz^{-1}} < L_{\nu} < 10^{26} \mathrm{WHz^{-1}}$.
This luminosity at which the model agrees with the observations decreases slightly with increasing redshift. The observations also have some significant uncertainties, for example at $L_{\nu} \sim 10^{25} \mathrm{WHz^{-1}}$ at $z=4$, the \cite{padovani15} errors are about 1.3 dex, while the observed number densities from \cite{simpson12} and \cite{padovani15} are different by 1 dex. Some of this tension between the model and the observations at higher redshifts may be a result of the observations wrongly attributing radio emission from star formation to radio emission from AGNs. 

In Figure \ref{fig:L800_radio_lf_evolution} we also show the evolution of the contributions to the luminosity function from hot halo mode accretion and starbursts triggered by mergers and disc instabilities. At $z=1$, similarly to $z=0$, the hot halo mode contribution dominates the luminosity function for low luminosities ($L_{\nu} < 10^{24} \mathrm{WHz^{-1}}$), the contribution from starbursts triggered by disc instabilities dominates for intermediate luminosities ($10^{24} \mathrm{WHz^{-1}} < L_{\nu} < 10^{25} \mathrm{WHz^{-1}}$), and the contribution from starbursts triggered by mergers dominates for high luminosities ($10^{25} \mathrm{WHz^{-1}} < L_{\nu} < 10^{27} \mathrm{WHz^{-1}}$). The hot halo mode contribution also dominates at the very highest luminosities ($L_{\nu} > 10^{27} \mathrm{WHz^{-1}}$), unlike at $z=0$. At $z=3$, the contribution from starbursts triggered by disc instabilties dominates at low luminosities ($L_{\nu} < 10^{25} \mathrm{WHz^{-1}}$) and the hot halo mode contribution dominates at higher luminosities ($L_{\nu} > 10^{25} \mathrm{WHz^{-1}}$). This behaviour continues out to $z=6$. 

In this figure, it can be seen that the model provides a slightly better fit to the observations for $z \geq 1$ with only the hot halo mode contribution. In \cite{griffin19a}, the inclusion of starbursts (particularly starbursts triggered by disc instabilities) gave better agreement between the model AGN radiative bolometric luminosity function and the observations for $0<z<6$. Here, we are exploring AGN properties from an existing galaxy formation model which includes disc instabilities to reproduce other galaxy properties \citep[see][]{lacey16}, but this tension regarding whether the model requires disc instabilities will require further investigation. 

\section{Evolution of the core radio luminosity function of AGN}
\label{sec:core_radio}

\begin{figure*}
\centering
\includegraphics[width=.8\linewidth]{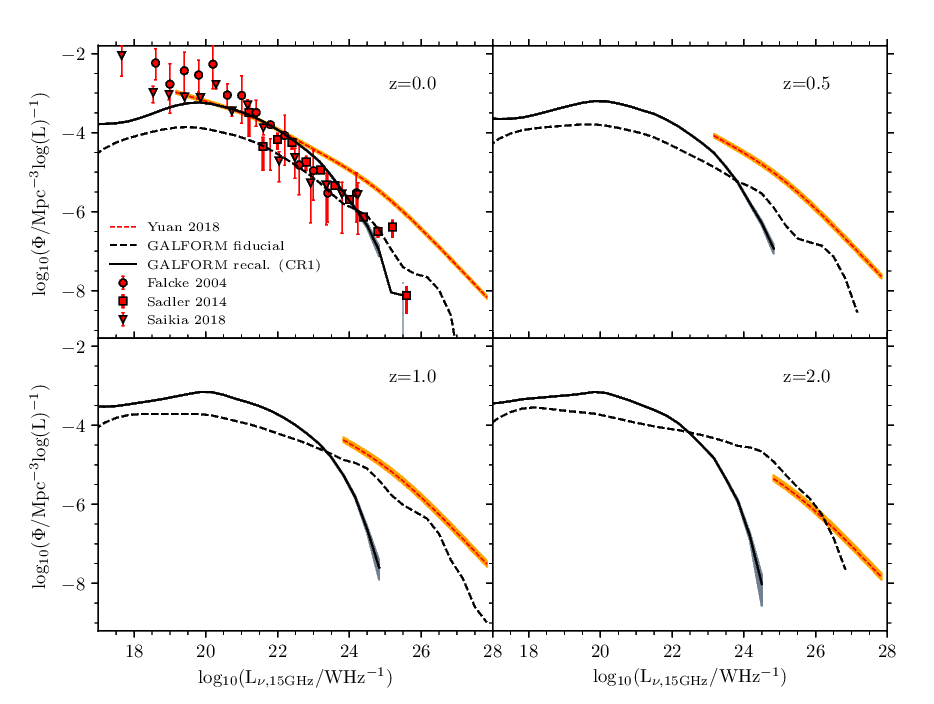}
\caption{The evolution of the rest-frame 15GHz AGN luminosity function predicted by model CR1 (with $A_{\mathrm{ADAF}}$ and $A_{\mathrm{TD}}$ given in Table \ref{tab:A_values}, shown as the black solid line with shading for errorbars) compared to observational estimates of the core radio luminosity function from \protect\cite{falcke04} (red circles), \protect\cite{sadler14} (red squares), \protect\cite{saikia18} (red triangles), and \protect\cite{yuan18} (red dashed line and orange shading). For this recalibration of the model, we require $A_{\mathrm{ADAF}}=A_{\mathrm{TD}}$. We also show the prediction from the fiducial model from Section \ref{sec:total_radio} (black dashed line), where we have converted to 15GHz luminosities using $L_{\nu} \propto \nu^{-0.7}$.}
\label{fig:core_radio_lf_evolution_same_As} 
\end{figure*}

\begin{figure*}
\centering
\includegraphics[width=.8\linewidth]{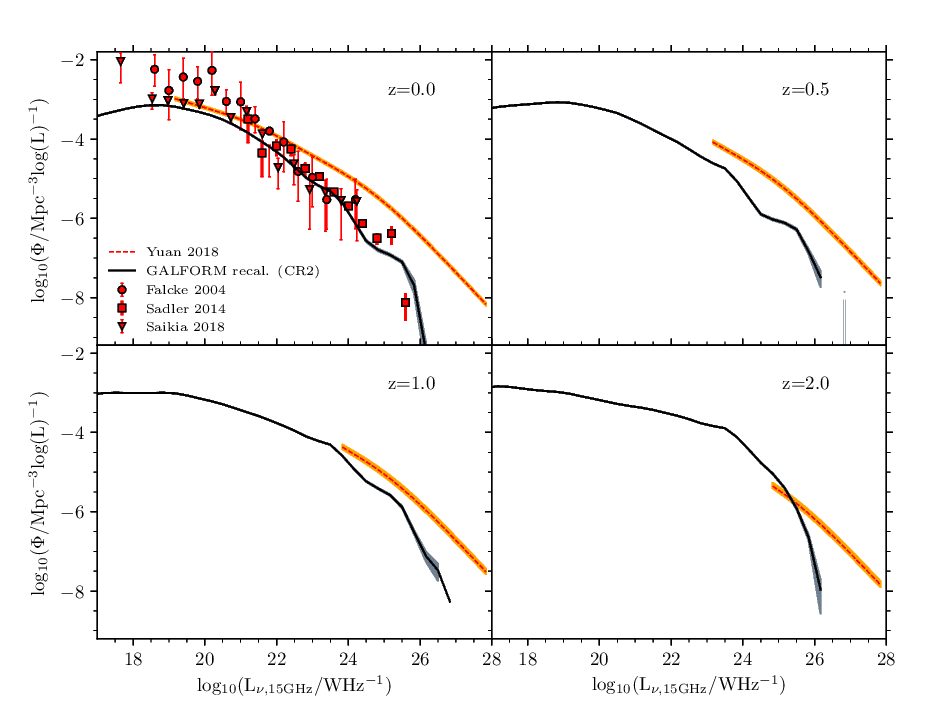}
\caption{The same as Figure \ref{fig:core_radio_lf_evolution_same_As}, but showing recalibrated model CR2 (with $A_{\mathrm{ADAF}}$ and $A_{\mathrm{TD}}$ given in Table \ref{tab:A_values}), in which we allowed $A_{\mathrm{ADAF}}$ and $A_{\mathrm{TD}}$ to be different.}
\label{fig:core_radio_lf_evolution} 
\end{figure*}

\begin{figure*}
\centering
\includegraphics[width=.8\linewidth]{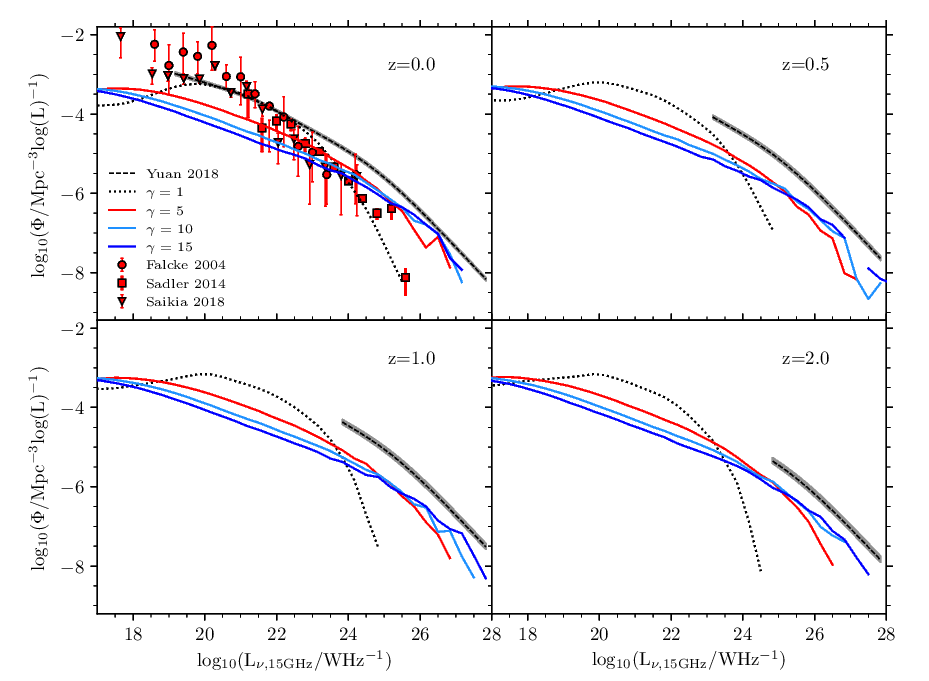}
\caption{The evolution of model CR1 including Doppler boosting for $\gamma=1$ (which corresponds to a Doppler boost factor $D=1$, black dotted line), $\gamma=5$ (red solid line), $\gamma=10$ (light blue solid line), and $\gamma=15$ (dark blue solid line). We compare the model to the same observational estimates of the core radio luminosity function in Figure \ref{fig:core_radio_lf_evolution_same_As}, except with the estimate of \protect\cite{yuan18} shown in black and grey for clarity.}
\label{fig:core_radio_lf_evolution_Dopp} 
\end{figure*}

AGNs show radio emission from both compact cores and extended components. The extended component generally dominates the radio emission at lower frequencies, whereas the core component is thought to originate from the inner jet, and dominates the radio emission at higher frequencies. In the radio luminosity functions presented in Figures \ref{fig:L800_radio_lf_z0} and \ref{fig:L800_radio_lf_evolution}, we calculated radio luminosities from equations (\ref{eq:lr_adaf}) and (\ref{eq:lr_td}) using the values of $A_{\mathrm{ADAF}}$ and $A_{\mathrm{TD}}$ from Table \ref{tab:free_params}, and we assumed that the radio luminosity is dominated by emission from the central core. We now consider the effect on our results if we compare our model instead to observational estimates of the luminosity function of core radio emission. To do this, we recalibrate the values of $A_{\mathrm{ADAF}}$ and $A_{\mathrm{TD}}$ to try to fit observational estimates of the luminosity function of core radio emission at $z=0$. In this Section, we present two recalibrations of the model, with the values of $A_{\mathrm{ADAF}}$ and $A_{\mathrm{TD}}$ given in Table \ref{tab:A_values}.

In \cite{falcke00}, \cite{sadler14}, and \cite{saikia18}, the luminosity functions of core radio emission are estimated from high frequency (15GHz or 20GHz) observations. At these high frequencies, the radio emission is expected to be dominated by the core emission rather than the extended emission. \cite{sadler14} find that for the majority of their sources, the radio emission is unresolved on $\sim 10 \mathrm{kpc}$ scales. \cite{sadler14} present a 20GHz radio luminosity function, which we assume is dominated by the core emission. On the other hand, \cite{falcke04} and \cite{saikia18} used high resolution Very Large Array (VLA) imaging to confirm the compact (parsec scale) natures of the radio cores in their samples. The contribution of radio emission from star formation is expected to be relatively small at these high frequencies.

A different study \citep{yuan18} estimated the core radio luminosity function by using an observed total radio luminosity function and a derived relation between core radio luminosity and total radio luminosity. Importantly, this relation is assumed to be independent of redshift. \cite{yuan18} only present their core luminosity function as a double power-law fit. As their estimates are more model-dependent than the others, we show their estimates as a dashed line to emphasise the difference between it and the other estimates. We only show the \cite{yuan18} estimate for $z \leq 2$, as beyond this their sample is very sparse, and also only plot the luminosity function over the core luminosity range directly observed in their sample\footnote{We use the luminosity range of objects in the sample of \cite{yuanwang12}, on which the sample used in \cite{yuan18} is based.}. 


We present core radio luminosity functions at 15GHz, converting the observations that are not at 15GHz \citep{sadler14,yuan18} assuming a flat spectrum ($\alpha=0$), which is appropriate for compact sources, and also assuming a flat spectrum for the model. We first explore whether a good fit to the observed core radio luminosity function can be obtained while requiring $A_{\mathrm{ADAF}}$ and $A_{\mathrm{TD}}$ to have the same value. With this requirement, we present model CR1, as shown in Figure \ref{fig:core_radio_lf_evolution_same_As}. We have primarily calibrated this model to the observational estimates from \cite{falcke04}, \cite{sadler14}, and \cite{saikia18}. The model is in good agreement with these three observational estimates at $z=0$, but underpredicts the estimate of \cite{yuan18} at high luminosities ($L_{\nu} > 10^{24}\mathrm{WHz^{-1}}$). The model underpredicts the estimate of \cite{yuan18} at higher redshifts. We therefore explore whether a better fit can be obtained by allowing $A_{\mathrm{ADAF}}$ and $A_{\mathrm{TD}}$ to be different. 

In Figure \ref{fig:core_radio_lf_evolution}, we present model CR2, where $A_{\mathrm{ADAF}}$ and $A_{\mathrm{TD}}$ have been allowed to be different. As for model CR1, we have primarily calibrated this model to the observational estimates from \cite{falcke04}, \cite{sadler14}, and \cite{saikia18}. This model fits slightly better to the observational estimates at $z=0$, and although it generally underpredicts the estimate of \cite{yuan18} for $z<2$, it is approximately consistent with \cite{yuan18} at $z=2$. At $z=2$, the model does predict a different shape for the core radio luminosity function than the observations, being steeper at high luminosities than the estimate of Yuan et al. This model is therefore more consistent with the observations than model CR1.

We also explored whether a better fit could be obtained by including Doppler boosting effects. The \cite{heinzsunyaev03} model supposedly takes into account the angle-average effect of Doppler boosting effects, but Doppler boosting depends strongly on the angle between the jet and the line of sight to the observer. Doppler boosting is where sources pointing towards the observer have a higher luminosity because of relativistic beaming effects. The luminosity of sources is changed by $L_{\mathrm{o}} / L_{\mathrm{i}} = D^{3+\alpha}$ where $L_{o}$ is the observed luminosity of the source, $L_{\mathrm{i}}$ is the intrinsic luminosity of the source when Doppler boosting is not included, $\alpha=0$ is the spectral index of the source, and $D$ is the Doppler factor, given by:

\begin{equation}
    D = \frac{1}{\gamma (1-\beta \cos\theta)},
\end{equation}

\noindent
where $\gamma$ is the bulk Lorentz factor of the jet, $\beta$ is the jet speed in units of $c$, and $\theta$ is the angle between the jet and the line of sight to the observer. We assume that the jet directions are randomly distributed over the surface of a sphere. Doppler boosting produces a tail of objects with large boost factors, but causes most objects to have lower luminosities. The distribution of $D^{3+\alpha}$ is a power law with slope $- 1/(3+\alpha)$, with minimum and maximum values that depend on $\gamma$, where higher values of $\gamma$ have a wider range of values. The angular dependence of optical depth effects is not included here, which may be important. We might expect that sources with the strongest Doppler boosting would have the largest optical depth.

In Figure \ref{fig:core_radio_lf_evolution_Dopp} we then compare the model prediction including Doppler boosting to observations of the core radio luminosity function. We show $\gamma=1,5,10,15$ \citep[similar to the values estimated by][]{ghisellini10}, where $\gamma=1$ corresponds to $D=1$. We use the same parameters as for model CR1 to illustrate the effect of including Doppler boosting. The model including Doppler boosting fits less well to the observational estimates by \cite{falcke04}, \cite{sadler14}, and \cite{saikia18} at both lower luminosities ($L_{\nu} < 10^{22}\mathrm{WHz^{-1}}$) and higher luminosities ($L_{\nu} > 10^{24}\mathrm{WHz^{-1}}$) at $z=0$, but is in better agreement with the observational estimate of \cite{yuan18} at the highest luminosities ($L_{\nu} \sim 10^{26}\mathrm{WHz^{-1}}$) at $z=0$. While the model predicts fewer objects than \cite{yuan18} at all redshifts, the slope of the luminosity function is similar at $z=2$.

The assumption in \cite{yuan18} that the relation between core and total radio luminosity does not evolve with redshift may be the cause of the discrepancy between model CR1 and CR2 and the observations at higher redshift (when including Doppler boosting the model is in better agreement with the observations at higher redshift). Alternatively the discrepancies may arise because the simple model we are using for the core-dominated emission is inaccurate.  

\vfill

\section{Conclusions}
\label{sec:conclusions}

Understanding the evolution of radio AGN is important for understanding galaxy evolution, given the role that relativistic jets from AGNs are believed to play in shutting off star formation via AGN feedback. Observational estimates of the evolution of the radio luminosity function have greatly improved in recent years, and these can provide insights into the evolution of AGN jets. 

We present predictions from the \galform semi-analytical model of galaxy formation, where a merger tree describing the formation history of each dark matter halo is populated with galaxies using analytical prescriptions for the baryonic physics. The model tracks black hole mass and spin evolution, where black holes can grow by quiescent accretion of gas from the hot halo, or by accretion of gas from starbursts triggered either by galaxy mergers or disc instabilities, or by merging with another SMBH. The gas accretion and the mergers also change the SMBH spins. We calculate jet luminosities using a Blandford-Znajek type model, and calculate the radio luminosity from the jet power using the \cite{heinzsunyaev03} scaling model.

First, we predict the evolution of the jet powers. We present the distribution of jet powers of objects for $0 \leq z \leq 6$, finding that the hot halo mode and ADAF accretion state contributions dominate for higher jet powers ($Q \gtrsim 10^{33}$) at lower redshift ($z=0$), contribute approximately equally to the starburst mode and thin disc accretion state at $Q \sim 10^{36}\mathrm{W}$ at $z=3$, but do not dominate the jet power distribution at the highest redshifts ($z=6$). The starburst mode and thin disc accretion state contributions dominate at low jet powers and low redshift, and at all jet powers at higher redshifts. We find that the peak of this distribution, which dominates the jet power density, is at $Q \sim 10^{36}\mathrm{W}$, independent of redshift for $0<z<6$. The distribution for the starburst and hot halo contributions also peaks at this jet power.

We then explore the predicted evolution of the jet power density. The jet power density is dominated by the contribution from haloes of mass $13 < \log (M_{\mathrm{halo}} / M_{\odot})  < 14$ for $z<1$, and by the contribution from haloes of mass $11 < \log (M_{\mathrm{halo}} / M_{\odot}) < 13$ for $z>1$. The mean AGN jet efficiency, which is the ratio of the jet power density to the mass accretion rate density, for the hot halo mode only varies modestly with time, suggesting that the assumption in the galaxy formation model that AGN heating efficiency is constant through time is a reasonable assumption. We then compare the jet power density evolution to the observational estimate of \cite{ceraj18} based on the measured radio luminosity function. The model prediction is slightly higher than the observational estimate, but reproduces the general shape of the jet power density evolution. The model evolves somewhat less steeply than the observations at low and high redshifts. Given the uncertainties in observationally estimating jet powers from radio luminosities, this tension may not be significant. 

We then present the predicted radio luminosity function, where the two free parameters of the model relating radio luminosity to jet power ($A_{\mathrm{ADAF}}$ and $A_{\mathrm{TD}}$) are calibrated to the observed total AGN radio luminosity function (including both compact and extended radio emission) at $z=0$. We split the predicted radio luminosity function at $z=0$ into contributions from different gas fuelling modes, finding that the contribution from the hot halo mode dominates at low luminosities ($L_{\nu} < 10^{24} \mathrm{WHz^{-1}}$), the contribution from starbursts triggered by disc instabilities dominates at intermediate luminosities ($10^{24} \mathrm{WHz^{-1}} < L_{\nu} < 10^{26} \mathrm{WHz^{-1}}$), and the contribution from starbursts triggered by mergers dominates at high luminosities ($L_{\nu} > 10^{26} \mathrm{WHz^{-1}}$). We also find that the radio luminosity function at $z=0$ at intermediate luminosities ($10^{22.5} \mathrm{WHz^{-1}} < L_{\nu} < 10^{26.5} \mathrm{WHz^{-1}}$) is dominated by the contribution from AGNs in haloes of mass $11 < \log (M_{\mathrm{halo}} / M_{\odot})  < 13$, whereas at $L_{\nu} < 10^{22.5} \mathrm{WHz^{-1}}$ and $L_{\nu} > 10^{26.5} \mathrm{WHz^{-1}}$, the radio luminosity function is dominated by AGNs in haloes of mass $13 < \log (M_{\mathrm{halo}} / M_{\odot})  < 15$. 

We present predictions for the evolution of the radio luminosity function in the redshift range $0<z<6$. The predictions evolve similarly to the observations of the total radio luminosity function, although at higher redshift the model luminosity function is steeper than that implied by observations. At the highest redshifts ($z>3$) we find that the radio luminosity function is dominated by the contribution from starbursts triggered by disc instabilities for $L_{\nu} < 10^{25} \mathrm{WHz^{-1}}$ and by the contribution from the hot halo mode for $L_{\nu} > 10^{25} \mathrm{WHz^{-1}}$. 

Finally, we recalibrated the model for comparison with the luminosity function for core radio emission in AGNs. First, we recalibrated the model with the constraint that the two free parameters that relate radio luminosity to jet power ($A_{\mathrm{ADAF}}$ and $A_{\mathrm{TD}}$) must be equal. This model gave an acceptable fit with most observations at $z=0$, but not very good agreement for $0.5 \leq z \leq 2$. Slightly better agreement at all redshifts was obtained when we allowed $A_{\mathrm{ADAF}}$ and $A_{\mathrm{TD}}$ to be different. We also present predictions where radio sources in the model have been Doppler boosted, which gives better agreement with the slope of the observed core radio luminosity function at $z=2$.

While the model generally provides a reasonable fit to the observational data, and incorporates some key physics by calculating jet powers and radio luminosities using a prescription based on SMBH mass, accretion rate and spin,
it could be improved. The scaling model that we use to relate the radio luminosity to the jet power was developed for core-dominated radio emission, however, 
in real radio AGN there is also radio emission from extended lobe structures, 
which are particularly important at lower frequencies. Inclusion of a more detailed model for the radio emission from AGN including extended emission from lobes might provide a better fit to the observations (e.g. the slope of the 
high redshift luminosity function), and so we plan to present
such a model in a future paper. 

Overall, the comparison of predictions like these to observations is important for an improved understanding of AGN feedback, and here we provide a first step by presenting the evolution of jet powers and radio luminosities from our model. Further development of this model should allow more refined comparisons with observational data, and so provide more detailed insight into AGN feedback.

\section*{Acknowledgements}

We thank Vernesa Smol\v{c}i\'c and Lana Ceraj for useful discussions, and thank Philip Best and Richard Bower for helpful comments and suggestions. This work was supported by the Science and Technology facilities Council grants ST/L00075X/1 and ST/P000541/1.
AJG acknowledges an STFC studentship funded by STFC grant ST/N50404X/1. 
This work used the DiRAC Data Centric system at Durham University, operated by
the Institute for Computational Cosmology on behalf of the STFC DiRAC HPC
Facility (www.dirac.ac.uk). This equipment was funded by BIS National
E-infrastructure capital grant ST/K00042X/1, STFC capital grants ST/H008519/1
and ST/K00087X/1, STFC DiRAC Operations grant ST/K003267/1 and Durham
University. DiRAC is part of the National E-Infrastructure.




\bibliographystyle{mnras}
\bibliography{references}







\bsp	
\label{lastpage}
\end{document}